# Probing Model Signal-Awareness via Prediction-Preserving Input Minimization

Sahil Suneja[*]
IBM Research
Yorktown Heights, NY, USA
suneja@us.ibm.com

Yunhui Zheng[*]
IBM Research
Yorktown Heights, NY, USA
zhengyu@us.ibm.com

Yufan Zhuang[*]
IBM Research
Yorktown Heights, NY, USA
yufan.zhuang@ibm.com

Jim A. Laredo
IBM Research
Yorktown Heights, NY, USA
laredoj@us.ibm.com

Alessandro Morari
IBM Research
Yorktown Heights, NY, USA
amorari@us.ibm.com

## ABSTRACT

This work explores the signal awareness of AI models for source code understanding. Using a software vulnerability detection use case, we evaluate the models' ability to capture the correct vulnerability signals to produce their predictions. Our prediction-preserving input minimization (P2IM) approach systematically reduces the original source code to a minimal snippet which a model needs to maintain its prediction. The model's reliance on incorrect signals is then uncovered when the vulnerability in the original code is missing in the minimal snippet, both of which the model however predicts as being vulnerable. We measure the signal awareness of models using a new metric we propose- Signal-aware Recall (SAR). We apply P2IM on three different neural network architectures across multiple datasets. The results show a sharp drop in the model's Recall from the high 90s to sub-60s with the new metric, highlighting that the models are presumably picking up a lot of noise or dataset nuances while learning their vulnerability detection logic. Although the drop in model performance may be perceived as an adversarial attack, but this isn't P2IM's objective. The idea is rather to uncover the signal-awareness of a black-box model in a data-driven manner via controlled queries. SAR's purpose is to measure the impact of task-agnostic model training, and not to suggest a shortcoming in the Recall metric. The expectation, in fact, is for SAR to match Recall in the ideal scenario where the model truly captures task-specific signals.

## 1 INTRODUCTION

Recently, AI research has made inroads in source code understanding and being able to perform tasks such as variable naming, function naming, summary generation, variable misuse detection, and vulnerability detection, amongst others [2–5, 22, 34, 44]. Ever more sophisticated models are emerging and pushing the state of the art rapidly. Although each new model improves upon its predecessor's prediction performance in terms of F1 and accuracy measures, what remains relatively unexplored is whether the models are picking up the *right* signals to arrive at their predictions.

We call this aspect of the model's quality 'signal awareness', and explore it using a software vulnerability detection use-case. What used to be a domain traditionally dominated by static and dynamic analysis is seeing assistance and competition from AI models. The high false positives of static analyzers, and the lack of completeness of dynamic analysis are a few reasons promoting the entry of AI into this field [27, 56, 62]. However, unlike the rules and path/flow analysis of static analyzers and the execution tracing of dynamic analysis, it remains unclear as to what signals the AI models actually pick up for detecting vulnerabilities in source code. From a model's perspective, it might be doing an excellent job of learning a separator between healthy and buggy code samples. But it may very well be the case that it does so by picking up noise or certain nuances from the dataset for their predictions, which are not representative or even related to vulnerabilities. Although this can lead to high accuracy scores for AI-based detectors, which may be perfectly acceptable in a theoretical or statistical setting, this can lead to a false sense of security when applied to real-world usage as substitutes for traditional source code analyzers.

To this end, in this paper we present a systematic approach to uncover a model's vulnerability detection *logic* and evaluate its ability to capture real signals. We use a data-driven explainability approach akin to how the model would have developed its data-driven logic. We borrow a fault isolation technique from the Software Engineering domain called Delta Debugging [64]. The core idea of our *Prediction-Preserving Input Minimization* (P2IM) approach is to first reduce the original source code input to a trained model into a *minimal* snippet, without changing the model prediction. And then to verify whether the minimal snippet has the same vulnerability profile as the original code. The model's reliance on incorrect signals can then be uncovered when a vulnerability in the original code is missing in the minimal snippet, both of which the model however predicts as being vulnerable. Additionally, we present a new metric called *Signal-aware Recall (SAR)* to measure how well a model captures task-specific signals.

We apply P2IM on three different neural network architectures, operating at different popular representations of source code in the AI domain: (i) a convolutional neural network (CNN) treating code as a photo, (ii) a recurrent neural network (RNN) treating code as a linear sequence of tokens, and (iii) a graph neural network (GNN) operating on code as a graph. We apply these models for vulnerability detection over three different datasets, including a real-world dataset we extract from Github. Results show more than 85% of samples can be reduced while maintaining their vulnerable prediction, across datasets and models. However, we observe a sharp performance drop across all models when probing them for signal awareness, with Recall dropping from the high 90s to low 60s with our new metric, suggesting that the models are presumably picking up features irrelevant to the task at hand.

---
[*]Authors contributed equally to this research.

Our work calls for a signal-aware supplement to the traditional statistical measurements for AI models and efforts to focus more on the *real* signals during model training. *Note that we are not suggesting a shortcoming in the Recall metric. In fact, the expectation is for the model's SAR to reach its Recall in the ideal scenario where the model truly captured task-specific signals.* We strongly believe that the models are powerful enough to pick such relevant signals, with appropriate *guidance* during training. But to motivate research in that direction, the shortcomings of current task-agnostic training need to be revealed, which is precisely what we tackle in this work.

In our P2IM approach, to probe a model's signal awareness, we maintain utmost fairness to the model. We keep the model untouched, and never alter its training process or change its training distribution at all, which is performed on the original unmodified dataset. The model's Recall and SAR are evaluated on the dataset's original test-set itself. Note that this work does not identify *what* a model is learning, but rather *if* the model is learning vulnerability-related signals. We treat the trained model as a black box and query it for its prediction on iteratively smaller versions for each vulnerable sample from the test-set. With such input "perturbations", our goal is not to craft programs for adversarial attacks [6, 14, 54, 60], but to uncover the signal-awareness of a black-box model in a data-driven manner via controlled queries. That is, at a very superficial level, we ask the trained model if it feels a code sample 'AAAA' is vulnerable, where 'A' represents any atomic code chunk granularity; then if it feels 'AAA' is vulnerable, then if it feels 'AA' is vulnerable, and so on until it's verdict changes. We then test if the *minimal* reduced version, which preserves the model's verdict, actually contains the original ground-truth vulnerability. In our experiments, we use the Infer tool [12] to verify bug existence in the reduced samples. However, P2IM is independent of the specific bug-checker being employed. To ensure model fairness, we only feed valid compilable reduced samples to the model for its verdict, without introducing any new bugs.

Our method augments the common metrics toolbox with an alternative way to examine the model quality, giving reliability and trustworthiness to black-box AI models. P2IM brings in multiple enhancements as compared to popular perturbation-based model interpretability methods that work on individual input instances, such as LIME [43] and others as summarized in a recent survey [10]. While other approaches are able to derive parts of an input that contribute most to the model's final prediction, unlike our method, they cannot tell if the highlighted parts are the correct task-relevant signals. Another contrasting capability P2IM offers is to quantify how well a model learns the correct signals, thereby providing a signal-aware mechanism to compare different models. This is especially useful when competing models have comparable performance on traditional metrics (e.g., F1). Furthermore, the search space, and thus the time complexity, for such approaches can be huge, given the possible combinations of the different parts of the inputs to be explored. Thanks to the Delta Debugging algorithm, P2IM directly minimizes the input, significantly accelerating the search for the relevant parts of the input.

The salient features of our prediction-preserving input minimization approach (P2IM) are summarized as follows:
- **Black-box:** P2IM requires no knowledge about the target model's internals, and is applicable to all model types, including classic machine learning as well as neural networks.
- **Externally Verifiable:** Emitted minimal code is valid and compilable, enabling cross-verification with traditional static and dynamic analysis tools.
- **New Metric:** We present a new Signal-aware Recall metric to measure how well a model captures task-specific signals. This enables a more fair model evaluation and meaningful comparison.

This paper is organized as follows. We first discuss our motivation behind this work in Section 2, and then present a brief background on AI models for source code, as well as the Delta Debugging technique, in Section 3. Then, we present the details of our P2IM approach in Section 4, along with a few examples demonstrating P2IM in action. We evaluate the models' signal awareness using our approach in Section 5. Section 6 covers related work, and finally Section 7 concludes this paper.

## 2 MOTIVATION

The rapid proliferation of AI for source code understanding is leading to ever more sophisticated models, which are getting bigger and better with each successive iteration. We have been experimenting with the state of the art to assess its quality from a practical perspective. We noticed that the models seem to suffer from weak generality and robustness–known AI frailties. These concerns become ever more important if the models are to be applied to sensitive tasks, such as ensuring code security. We observed the same weakness across several different models and datasets in a vulnerability detection setting, which led us to doubt the quality of the models in terms of *what* is it that they are actually learning. We thus started exploring ways to probe the models' *signal-awareness*.

Signal-awareness is different than correctness — a model can learn a perfect separator between buggy and healthy code, but it can very well arrive at the separator by picking up dataset nuances, as opposed to real vulnerability signals. This can be caused when, for example, the model picks up unexpected correlations between code samples and sample lengths, or variable names, or certain programming constructs, which may happen to differ for buggy and healthy samples in a particular dataset. Learning a separator based on these non-representative signals, which may lead to great-looking performance numbers, is perfectly acceptable from a model's perspective. The model is indeed doing its job of learning to classify, but this provides a false sense of security.

Signal-awareness or verifying if the models are learning the correct logic relevant to code analysis is crucial to generate trust in models if they are to be put into the field in competition to, or alongside, traditional static and dynamic analyzers. Furthermore, it adds an important measure of model quality, beyond the traditional statistical analysis measures, which can more fairly compare and guide improvements across model evolutions. In this work, we uncover this signal-awareness aspect of AI-for-code models and quantify how much impact it has on their robustness as well as reported performance numbers.

A typical example of weak robustness is when an image classifier's verdict on an input image changes on adding minor noise to the image imperceptible to the naked eye [6, 14, 54]. As shown in Table 1, we observe the same issue with AI-for-code vulnerability detection models, where even a 99 F1 model flips its prediction on *only slightly* different code variants. However, it should be reasonable to expect a high-quality model, which correctly picks up the real vulnerability signals, to demonstrate prediction robustness, if it is ever to be trusted in a practical security setting. Although such "perturbations" *can* be taken to be an adversarial attack on the model [60], and similarly there can be defenses against such attacks such as training the model with several different code variants. But, this line of thought is complementary to our work. These observations merely triggered our suspicion around a disconnect between the reported model performance numbers versus the actual task-aware learning, similar in spirit to other prevailing doubts regarding model quality [1, 7, 23].

Our goal is not to discredit AI-for-code models by reiterating their brittleness, but to uncover and quantify how much impact task-agnostic training has on their reported performance numbers. In particular, we evaluate the trained models on the original test-set itself, while querying the model with iteratively reduced versions of each sample, to uncover precisely what portion of the original sample does the model consider to be relevant for its prediction. Then, by counting the occurrences across the test-set, where the *minimal* portions still contain the same vulnerability as the original samples, we can measure how signal-aware the model is using our proposed SAR metric.

Our hope is that revealing these shortcomings would motivate future research towards more task-aware learning, potentially guided by the SAR-Recall divide, to better utilize the potential of AI for source code understanding.

## 3 BACKGROUND

In this section, we first briefly describe three different neural network models which have been popularly employed for learning over source code, each operating at a different code representation. These include (i) a convolutional neural network treating code as a photo, (ii) a recurrent neural network treating code as a linear sequence of tokens, and (iii) a graph neural network operating on code as a graph. In section 5, we shall evaluate these models on our proposed SAR metric. After the model descriptions, we cover the basics of the Delta Debugging technique [64] which we use to reduce the input samples to measure the models' signal-awareness.

### 3.1 AI-for-code Models

**Convolutional neural networks (CNNs)** learn on image inputs. These are made up of convolutional and pooling layers. The former act as filters to extract features from input images, learning increasingly complex patterns when the neural network becomes progressively deeper. Pooling layers, on the other hand, down-sample the features so as to intensify the signal and control the size of the neural network. This is done by selecting the most strongly activated neurons (i.e., max-pooling) or taking their average (i.e., mean-pooling). CNNs have been successfully employed in computer vision tasks such as image recognition and object detection [19, 28, 42, 51]. These have also been used in the context of vulnerability detection with slight modifications [44]. In such a setting, the source code tokens are first projected into an embedding space and then fed into a CNN as a real-value matrix, similar to an image.

**Recurrent neural networks (RNNs)** are designed to learn over sequential inputs, such as text and audio [35, 45]. A 'working memory' is maintained and updated by a series of input, output, and forget 'gates', using the current input and the previous memory at each step [9, 21]. Depending on the use case, RNNs can emit vector representations at each step, or emit one final representation for a complete sequence. RNNs have been applied to source code by treating it as sequences of tokens. Usually, one final representation is extracted for a piece of code in the context of vulnerability detection [34, 44].

**Graphical neural networks (GNNs)** have gained popularity due to their unique ability to learn over graph-structured data like social network graphs and molecular structures [13, 65]. Most GNNs are made up of three modules: (i) message passing, which decides how information is exchanged among nodes via edges, (ii) message aggregation, which determines how each node combines the received messages, and (iii) message updating, which controls how each node updates their representation after one cycle of information propagation [13, 26, 32]. Using GNNs on source code is a natural fit since multiple forms of graphs can be constructed on top of source code, such as abstract syntax tree, data flow graph, and control flow graph. They have achieved state-of-the-art performance on multiple

Table 1: Bad Robustness: a 99 F1 GNN model flips on only slightly syntactically different but semantically identical function variants. Dataset from [47], model from [53]. In this paper, we focus on type (a) reductions only.

**(a) Removing tokens changes model predictions**

```
01. int a = 99;            01. int a = 99;
02. int b = 57;            02. int b = 57;
03. char arr[69];          03. char arr[69];
04. if (b < a) {           04. if (b) {
05.   b = 78;              05.   b = 78;
06. }                      06. }
07. arr[b] = 'X';          07. arr[b] = 'X';
```

| | | |
|---|---|---|
| Ground Truth | Buggy | Buggy |
| Prediction | Buggy | Non-buggy |

**(b) Adding tokens changes model predictions**

```
1. int a = 93;             1. int a = 93;
2. char arr[55];           2. char arr[55];
3. arr[a] = 'X';           3. arr[a] = 'X';
4. return 0;               4. a = 43;
                           5. return 0;
```

| | | |
|---|---|---|
| Ground Truth | Buggy | Buggy |
| Prediction | Buggy | Non-buggy |

software engineering tasks, including vulnerability detection [68] and code summarization [30].

## 3.2 Delta Debugging

Delta Debugging (DD) [64] was first introduced to minimize failure-inducing bug reports for the Mozilla browser. Many of these bug reports contain long HTML files or user actions with lots of irrelevant information, which makes it challenging to understand the root causes. Therefore, developers were looking for techniques that can simplify test cases and generate minimized ones triggering the same failures. As a result, DD significantly simplified the crash-inducing inputs in an automated manner and hence enabled a productive bug diagnosis and repair experience.

The input to DD is a sequence satisfying some predefined oracle. For example, it's failure-inducing HTML files or user actions in the Mozilla web browser case. The goal is to find a subset of the input satisfying the following two requirements: (1) the subset leads to the same outcome; and (2) not a single element can be removed to preserve the outcome. Such a subset is called *1-minimal*.

---

**Algorithm 1** Simplified Delta Debugging Algorithm

**Input:**
  $T$: Oracle function. $T(x)$ is TRUE if $x$ has certain predefined property.
  $S$: Input sequence, where $T(S)$ = TRUE
**Output:**
  $S'$: Reduced outcome-preserving *1-minimal* sequence

1: **function** DD($T, S$)
2:   $n \leftarrow 2, S' \leftarrow S$
3:   Divide $S$ equally into $\Delta_1, ..., \Delta_n$ and the complements
     $\nabla_k = S - \Delta_k$, where $k = 1, ..., n$
4:   Test each $T(\Delta_1), ..., T(\Delta_n)$ and $T(\nabla_1), ..., T(\nabla_n)$
5:   **if** all FALSE **then**
6:     $n \leftarrow 2 * n$
7:     **if** $n > |S|$ **then**
8:       **return** $S'$
9:     **else**
10:      **go to** line 3
11:  **else if** $T(\Delta_i)$ = TRUE **then**
12:    $S' \leftarrow \Delta_i$
13:    $n \leftarrow 2, S \leftarrow \Delta_i$
14:    **if** $|S'| = 1$ **then**
15:      **return** $S'$
16:    **else**
17:      **go to** line 3
18:  **else if** $T(\nabla_j)$ = TRUE **then**
19:    $S' \leftarrow \nabla_j$
20:    $n \leftarrow n - 1, S \leftarrow \nabla_j$
21:    **go to** line 3

---

As presented in Algorithm 1, given an input sequence $S$ and an oracle function $T$, DD iteratively splits the input sequence and produces 1-minimal $S'$ in four main steps:

- *Split and test*. In each iteration, DD splits the sequence in consideration into $n$ segments and $n$ corresponding complements (line 3), where a complement is defined as $\nabla_i = S - \Delta_i$. DD tests all partitions using the provided oracle function $T$ (line 4) and checks if some partitions lead to the same outcome.
- *Reduce a subset*. If the test result of a subset $\Delta_i$ is the same as $S$, DD treats $\Delta_i$ as the sequence for the next iteration and resets the granularity $n$ (lines 11 – 17).
- *Reduce a complement*. Otherwise, if complement $\nabla_j$ is an outcome-preserving input (line 18), DD adjusts $n$ and explores it with the same granularity (line 20).
- *Operate on a finer granularity*. If none of the partition can preserve the outcome (line 5), DD doubles the partition number $n$ to split the sequence into smaller segments (line 6).

In each round, DD tries to reduce the scope to a subset. In the best-case scenario, DD works like a binary search, which can systematically and efficiently identify the *1-minimal*.

Our intuition behind data-driven model probing comes from this failure-inducing input simplification idea. Specifically, we build our method atop DD while replacing the Mozilla target with the prediction model, and the failure-inducing HTML files with vulnerable program samples. DD's process of identifying a minimal subsequence of the input which leads to the same output, then translates to identifying the minimal sub-program (1-minimal) which preserves the model's prediction. The model's signal-awareness is then determined by testing the 1-minimal for the original vulnerability existence.

## 4 DESIGN

Programming defects are an inevitable reality in software creation. Vulnerabilities arise when such defects fall in a security-related subset such as null pointer dereference, buffer overflow, use-after-free, amongst others. Static analyzers detect these vulnerabilities either by reasoning about the possible execution behaviours over a program model, or by matching defect-specific rules. Dynamic analysis, on the other hand, directly executes the program, exploring different execution paths to concretely expose the defects. Unlike the traditional analyzers, the logic of AI-for-code models is implicit, and not directly perceptible. In this section, we present our approach towards understanding this logic, while treating the models as black box entities. Given explaining *what* an AI model is learning is still an open problem, especially in the context of source code understanding, we frame our exploration in terms of detecting *if* the models are learning the vulnerability-relevant signals.

### 4.1 P2IM Workflow

Figure 1 depicts the overall flow behind our prediction-preserving input minimization (P2IM) approach. The sequence of operations is as follows:

**Step 1.** P2IM takes as input a trained model and a program sample which the model predicts to be vulnerable. This sample comes from the test-set of the dataset used for model training itself.

**Steps 2-4.** P2IM then iteratively keeps reducing the sample and querying the model for its prediction on the reduced subprogram, so long as the model maintains its vulnerability prediction. This process continues till a minimal snippet, called 1-minimal, is extracted from the program sample, such that removing even a single token from it would change the model prediction. To efficiently and systematically reduce a program sample, we employ the popular

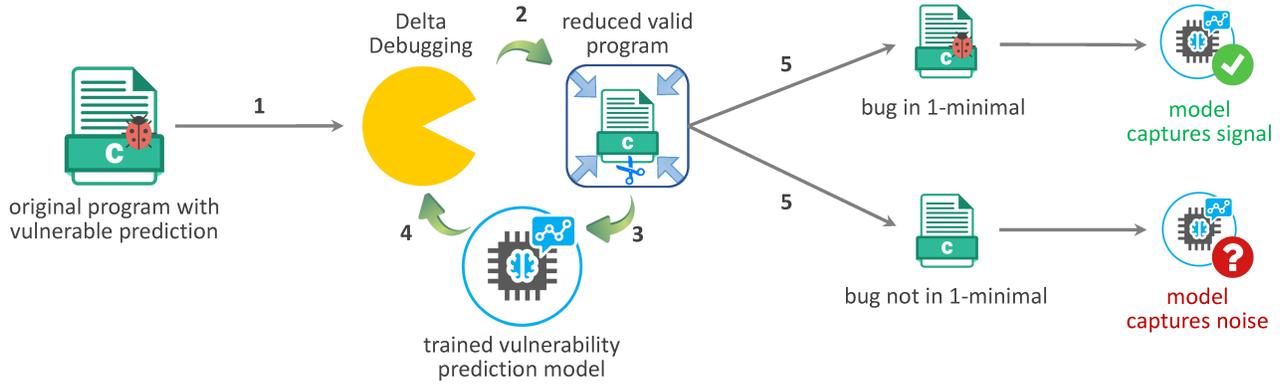

Figure 1: The end-to-end flow of P2IM approach: 1. A program sample, which the model correctly predicts to be vulnerable, is fed to DD. 2. DD produces a valid reduced sample. 3. The reduced sample is fed to the model for prediction. 4. Subsequent DD iterations are continued till 1-Minimal is found. 5. To test the signal awareness of the model, the 1-minimal is verified for the original sample's bug existence

---

**Algorithm 2** Prediction Preserving DD Oracle

**Input:**
  $S$: C/C++ program token sequence
  $M$: A vulnerability detection model
  $v$: Original program's vulnerability info

**Output:**
TRUE: $S$ represents a valid but vulnerable program per $M$. It should be further checked for reduction opportunities.

1: **function** T($S$)
2:   $p \leftarrow$ TOKEN2PROGRAM($S$)
3:   **if** VALIDPROGRAM($p$) **and** M($p$) = VULNERABLE **then**
4:     **if** SAMEVULN($p, v$) **or** NOVULN($p$) **then**
5:       **return** TRUE
6:     **else**
7:       **return** FALSE
8:   **else**
9:     **return** FALSE

---

fault isolation technique of Delta Debugging (DD) [64]. We follow the same procedure as outlined in Section 3.2, wherein DD reduces the program sample at the level of source code tokens, iteratively updating its split granularity till a valid 1-minimal sub-program is extracted. As discussed in Section 3.2, DD requires an oracle which drives the reduction cycle. This oracle decides whether or not an intermediate reduced subprogram should be picked for subsequent reductions. In order to produce a valid and prediction-preserving 1-minimal subprogram, as shown in Algorithm 2, we customize the oracle $T$ to require the reduced subprograms satisfy the following properties (line 3 and line 4):

- *Vulnerable prediction*. A subprogram is selected for further processing only if it preserves the model's vulnerability prediction.
- *Valid program*. We also enforce that the reduced subprogram is valid and compilable. Although we may be able to reduce more tokens by dropping this requirement, it is unfair to abuse the model by throwing random code at it, since it has (hopefully) been trained on valid programs as inputs. This additionally enables cross-verification with traditional program analysis tools.
- *Vulnerability type*. Beyond just asserting the correctness of the reduced subprogram as above, it is also verified for either possessing the same bug location and type as the original sample, or being bug-free, contingent on the oracle's quality. This is done to maintain fairness to the model, by not querying it with subprograms which have any additional bugs introduced by the minimization process itself.

**Step 5.** The 1-minimal produced in the previous step is a valid subprogram that the model predicts to be vulnerable. It represents *the bare minimum excerpt of the input sample, which the model needs to arrive at and stick to its original prediction*. The requirements specified in line 4 of Algorithm 2 ensure that the 1-minimal either has the same original vulnerability or is vulnerability-free, without any new vulnerabilities being introduced by the minimization procedure. Then, if this minimal excerpt indeed contains the same vulnerability as the original sample, P2IM treats it as a sign of the model learning vulnerability-specific signals. On the other hand, if the corresponding vulnerability from the original sample is missing in the 1-minimal, this points towards the model capturing noise or features not relevant to vulnerabilities.

### 4.2 P2IM Examples

We present two examples to demonstrate how P2IM minimizes program snippets, produces valid 1-minimal, and determines if a model learns vulnerability related signals.

**Model learned real signals**. Figure 2 shows an example walk-through of P2IM. In the original program, a buffer overflow can be triggered in "buf[b]=1;" when b>=10. We list the key steps to show how P2IM gradually adjusts the granularity of editing and reduces the tokens. In iterations #2-3, the token sequence is split into 2 parts but none can be a valid program. Hence, in iteration #4-7, DD operates on a finer granularity and the sequence is split into 4 parts ($n$ is increased to 4; see Algorithm 1, line 6). Since P2IM cannot find any valid subprograms, it doubles the sequence split to 8 parts ($n$ = 8), and finds a valid subprogram (iteration #13). Since the model predicts it to be vulnerable, statement "`a+3;`" can be removed. After that, P2IM finds iterations #15, #34, and #46 are valid subprograms with vulnerable predictions so that more tokens can be reduced. Given there are no smaller valid subprograms, iteration #46 is the prediction preserving 1-minimal. More importantly, iteration #46 is indeed vulnerable so P2IM counts it in favor of the model correctly capturing vulnerability-related signals.

Figure 2: Walk-through of P2IM's minimization- an example of a good model that captured real signals. Details in Section 4.2.

Figure 3: Testing model's signal-awareness with P2IM. Example comparing an original code snippet (Juliet dataset) vs. its 1-minimal version. The model (CNN, 97 F1, Section 5) considers the 1-minimal as buggy even though it does not contain the actual bug (highlighted in red). This suggests the model incorrectly learns non-vulnerability signals.

**Model missed real signals**. Figure 3 shows an example of P2IM catching a model incorrectly learning non-vulnerability signals. In the original program, the sink variable '`data`' points to a buffer of size 10, while the size of the source variable '`source`' is 11. Therefore, a buffer overflow can occur at line 11. The model correctly predicts this program as being vulnerable. However, the model considers its 1-minimal as being vulnerable as well, which however is missing the culprit assignment operations. This means the model doesn't even need the statement where the buffer overflow actually occurs, to make the vulnerable prediction. P2IM regards this as the model not learning vulnerability-specific signals.

### 4.3 Signal Awareness Measurement Bounds

**1-minimal versus global minimum**. It is a possibility that the 1-minimal reached by P2IM is not the global minimum for the original program. And the global minimum may not contain the vulnerability-related signals as captured in a local 1-minimal. Since computing the global minimum is impractical as it requires an exponential number of tests, this precludes measuring a model's signal awareness precisely. Thus, we take the conservative approach of giving the benefit of doubt to the model, giving it credit for capturing vulnerability signals based on the 1-minimal reached, even when it may not actually be doing so (based on the global minimum). This results in an upper bound measurement of the model's signal awareness (SAR). Nevertheless, as shall be revealed in the next section, even measuring the upper bound itself is sufficient to highlight the problems of the models not picking up the correct signals during learning.

**Checker quality.** The signal awareness measurement is bounded by the quality of the checker used to verify bug existence in the reduced subprograms. This can be dataset dependent and can include (i) the original dataset labeler, (ii) a line-based bug matcher which gives the benefit of doubt to the model on partial matches, or (iii) a good static analyzer tuned towards high recall. For the experiment settings and the datasets considered in this paper (Section 5.1), the Infer analyzer [12] worked quite well, with fallback to line-based bug matching for samples with differing Infer verdict and the original bug. Although Infer as a checker is a fortunate fit given our target datasets, P2IM is not reliant on it. While Infer provides for a more accurate SAR bound, a similar lack of signal awareness in AI-for-code models is still detected, albeit with a looser bound, by replacing Infer with purely a line-based bug matcher (less accurate; more model favoring). Although using the original dataset labeler might be even more accurate, and expands the dataset and task applicability of P2IM, but it can be a harder task especially with human-in-the-loop kind of labelers. Finally, the existence of a perfect checker precludes the need for AI for code analysis. Yet, to introduce some accountability in today's AI-for-code models, we argue for at least a SAR-like sanity check.

## 5 EVALUATION

We use the following methodology for testing the signal awareness of vulnerability detection models. The trained models will assign labels to the code samples in the test set as either being vulnerable or not, based upon the predicted class probabilities. After prediction, all vulnerable samples in the test set fall under either True Positives (TP) or False Negatives (FN). We subject each TP predicted by the model to P2IM reduction. We first query the model for its prediction on each TP sample's 1-minimal version. Then we check the 1-minimal for the presence (TP') or absence (FN') of the original program sample's bug. This way we subdivide TP into signal-aware TP' and signal-agnostic FN'.

Operating atop vulnerable samples (TP+FN), *Recall* is the best metric to target, to fairly compare different models on the same dataset. This is because number of vulnerable samples = TP + FN = TP' + FN' + FN will be the same for all models for the same dataset. We present a new metric- Signal-aware Recall (SAR), to measure the signal-awareness of vulnerability detection models. So, while Recall = TP / (TP + FN), SAR is defined as TP' / (TP' + FN' + FN), where TP = TP' + FN'. Then we compare the two metrics for each model, to highlight how much of its reported Recall is attributable to task-relevant signal learning (reflected by SAR).

For scalability, we run P2IM reduction across multiple samples in parallel. As an example of the runtime cost, it takes {min:3, avg:119, max:884} seconds for the Github dataset samples, described next.

### 5.1 Datasets

P2IM's signal awareness measurement requires datasets with ground truth bug locations. Datasets from Draper [44] and Devign [68] are excluded because they do not specify bug locations. Samples from VulDeePecker [34] and SySeVR [33] are slices converted into linear sequences, not valid compilable code which models are trained upon and thus excluded from experiments. Therefore, we use the following three datasets which contain this granularity of line-level bug information. The `train:validate:test` split is kept at `80:10:10` for all experiments.

*5.1.1 Juliet.* The Juliet Test Suite [40] contains synthetic examples with different vulnerability types, designed for testing static analyzers. From its 64K test cases, we extract 118K functions, amongst which almost 35% are vulnerable. Samples tagged as 'bad', and with clear bug information as per `manifest.xml` file, are labeled as 1, while the ones with a 'good' tag are labeled as 0.

*5.1.2 s-bAbI .* The authors of s-bAbI [47] claim that the Juliet dataset is far too small and complex to use in learning to predict the labeled security defects. Their proposed s-bAbI synthetic dataset contains syntactically-valid C programs with non-trivial control flow, focusing solely on the buffer overflow vulnerability. We used the s-bAbI generator to create a balanced dataset of almost 475K functions. Samples with a 'UNSAFE' tag are labeled as 1, and those with 'SAFE' as 0.

*5.1.3 Github dataset.* Different from the synthetic s-bAbI and Juliet datasets, we also include a real-word dataset with bug location and bug type information, which we derive from the D2A dataset [66]. D2A is trace-level dataset built over multiple Github projects- OpenSSL, FFMpeg, HTTPD, Nginx and libtiff. It is generated by using differential analysis atop the Infer static analyzer outputs of consecutive versions before and after bug-fixing commits. From D2A's traces, we derive function-level samples. From each before-fix trace associated with a bug, we extract the functions specified by the reported bug locations and label them as 1. We also extract the corresponding functions in the after-fix trace and label those patched by the corresponding commit as 0. After deduplication, we have 6728 functions in total.

### 5.2 Models

We apply P2IM on three different neural network architectures which have been popularly employed for vulnerability detection. These operate upon different representations of source code.

**CNN:** This model treats source code as a photo and tries to learn the pictorial relationship between source code tokens and underlying bugs. Similar to [44], we apply token normalization before feeding data into the model. This involves normalizing the function names and variable names to fixed tokens such as `Func` and `Var`. We set the embedding layer dimension as 13, followed by a 2d-convolutional layer with input channel as 1, output channel as 512, and kernel size as (9, 13). The final prediction is generated by a 3-layer multilayer perceptron (MLP) with output dimensions being 64, 16, and 2.

**RNN:** This model treats code as a linear sequence of tokens and tries to detect bugs in source code using the temporal relationship between its tokens. We base our RNN implementation on [33]. The input function is also normalized during preprocessing, the same as the CNN model. We set the embedding layer dimension as 500, followed by a two-layer bi-directional GRU module with hidden size equals to 256, the final prediction is generated by a single-layer MLP.

Table 2: Comparing AI-for-code models using standard as well as proposed Signal-Aware Recall (SAR) metric. Note the drop in model quality (Recall → SAR) when probing it for signal awareness.

| Dataset | Model | Accuracy | F1 | Recall | SAR |
|---|---|---|---|---|---|
| s-bAbI | CNN | 97.5 | 97.4 | **95.4** | **58.4** |
| s-bAbI | RNN | 97.6 | 97.5 | **95.9** | **59.6** |
| s-bAbI | GNN | 99.6 | 99.6 | **99.7** | **62.7** |
| Juliet | CNN | 97.9 | 96.9 | **99.9** | **77** |
| Juliet | RNN | 98 | 97 | **98.6** | **76** |
| Juliet | GNN | 99.6 | 99.4 | **99.9** | **76.8** |
| Github | CNN | 60 | 58.9 | **54.9** | **46.2** |
| Github | RNN | 57.1 | 60 | **61.8** | **52.4** |
| Github | GNN | 59.3 | 64.2 | **69.8** | **60.8** |

Table 3: P2IM reduction stats.

| Dataset | Model | % Samples Reduced | Average Reduction % |
|---|---|---|---|
| s-bAbI | CNN | 100 | 41 |
| s-bAbI | RNN | 100 | 42.7 |
| s-bAbI | GNN | 100 | 41.9 |
| Juliet | CNN | 85.9 | 38.8 |
| Juliet | RNN | 85.8 | 38.9 |
| Juliet | GNN | 86.2 | 38.9 |
| Github | CNN | 87 | 45.9 |
| Github | RNN | 88.5 | 45.5 |
| Github | GNN | 87.1 | 43.6 |

**GNN:** Instead of borrowing techniques from image and time-series domain, this model operates at a more natural graph-level representation of source code, as per [53, 68]. It tries to learn vulnerability signatures in terms of relationships between nodes and edges of a Code Property Graph [59]. Following [68], we do not apply token normalization during preprocessing. We set the embedding size as 64, followed by a GGNN layer [32] with hidden size 256 and 5 unrolling time steps. The node representations are obtained via summation of all node tokens' embedding, and the graph representation read-out is constructed as a global attention layer. The final prediction is generated by a 2-layer MLP with output dimensions 256 and 2.

The models are trained over the datasets presented in Section 5.1, with the vulnerability detection problem framed as a binary classification task– predicting program samples as healthy (label 0) or buggy (label 1). For all of the models, we set dropout rate as 0.2 during training, and used the Adam [25] optimizer. We tuned learning rate in $\{10^{-3}, 10^{-4}\}$ and batch size in $\{24, 36, 128, 256, 512\}$. Models are trained with 100 maximum epochs and early stopping (patience = 10). Cross entropy loss, with class weight calculated from the training set, is employed. For each experiment, we save the checkpoint with the best validation loss.

### 5.3 Results

Table 2 compares the performance of the three models under test, upon the three datasets as described earlier. Included are the common measures of the models' classification performance. The model reproductions achieve performance similar to previous work [44, 53], including on the real-world dataset with performance comparable to that of [68], which also creates its dataset from Github but lacks bug location information.

The focus is on how Recall compares with the proposed SAR metric. As can be seen, even for the simple synthetic s-bAbI dataset, which targets only one vulnerability type (buffer overflow), a 95+ Recall drops into the sub-60 range across the board, when we probe the models for signal awareness with our P2IM approach. This indicates that the models are picking up features not relevant to vulnerability detection, presumably learning dataset nuances which inflates their reported performance measures. The results are similar for the other datasets as well.

Table 3 shows the reduction statistics obtained with P2IM. Column 4 shows that a significant reduction of 39%-46% can be achieved in the samples, without the models changing there prediction, and the rates are similar across the different models for the same dataset. Furthermore, more than 85% of the samples can be reduced while maintaining their vulnerable prediction, across datasets and models. The rest are a mix of two categories:

(1) The samples which cannot be reduced due to the valid code requirement which we enforce on P2IM, so as to only feed real compilable subprograms to the models. For these, we cannot ascertain for sure whether the model truly captures the vulnerability signals.
(2) The samples which the model truly needs as is for making its prediction, signifying a potential true signal capture.

We observed the compilable code requirement is the main reason why tokens cannot be reduced. Taking the example of CNN+Juliet, amongst the 14.1% of TPs (534/3775) which cannot be reduced, 89.1% of them (476/534) have Windows headers which are not compilable in our (Linux) environment. Similarly, for CNN+Github, all of the 13% TPs which cannot be reduced are due to compilation failures.

Table 4 shows the overlap between the subset of the program samples with the same prediction across different models. On the two synthetic datasets, the overlap percentage is >90% for the true positive samples (TP), as well as for the signal-aware true positives (TP'), across all three models. This is unexpected since the models use vastly different architectures. Combined with the low SAR values from Table 2, this suggests that the perfect performance on synthetic datasets is significantly influenced by dataset-specific nuances, which all three models are picking up in a very similar way, and missing real vulnerability signals. The corresponding overlap on the real-world dataset is much lower, partially due to the fact that there are more variety and less artificial nuances in real-world data for models to pick up, which however contributes to their performance drop when compared to synthetic datasets, as shown in Table 2.

Table 4: Overlap between the subset of samples with the same prediction across different models. Shown is the overlap percentage for the TP (as well as signal-aware TP') subsets across the {CNN, RNN, GNN} models for each dataset.

| Dataset | TP Overlap % | TP' Overlap % |
| --- | --- | --- |
| s-bAbI | 95.3 | 91 |
| Juliet | 97.4 | 96.9 |
| Github | 49 | 49.3 |

As a takeaway, with the addition of SAR to the existing arsenal of model performance metrics, it becomes possible to measure how much of the model's learning is actually task-aware. This can additionally provide a more fair comparison, and more accurate improvement guidance across model evolutions.

## 6 RELATED WORK

In the software engineering community, efforts have been made to detect vulnerabilities by isolating relevant statements using program slicing based techniques [18, 48, 55, 59]. Essentially, the main idea of such approaches is to identify the subsets of the program that introduce the defects. Recently, program slicing has also been used in AI-assisted vulnerability detection tasks [34] to extract bug relevant programming constructs. If a vulnerability detection model can capture the real signals, it should be able to identify such subsets too. In this work, we treat models as black boxes and feed them with different subsets of a program to evaluate how well they pick up the real signals. And to efficiently and systematically generate the subsets, we borrow a popular fault isolation technique of Delta Debugging.

The most relevant work using Delta Debugging (DD) and its variant methods is software failure diagnosis and isolation [17, 38, 63, 64]. The main advantage of DD is that it can significantly reduce the number of tests needed to locate the problem. To the best of our knowledge, our approach is the first attempt of using delta debugging to interpret and compare models' signal awareness. A recent parallel effort [41] also uses DD to minimize inputs to AI models on software engineering tasks. Its focus is on the qualitative properties of the reduced code samples, as opposed to our SAR-based quantification of the impact of signal-agnostic training on the models' reported performance numbers.

P2IM can be viewed as belonging to the metamorphic testing paradigm [8], applied to AI models [20, 69]. In particular, based on an input code snippet and its prediction by a model under test, we systematically construct new "tests" by minimizing the original snippet and make sure the model produces the same prediction as the original input. Then we check the metamorphic relations among the inputs and output predictions of multiple executions. A test violation can be detected if the original input and its minimal version do not have the same vulnerability.

In contrast to classical statistical learning models that have explicit input & output relationship and error bounds, it is hard to understand what deep learning models learn and to provide robustness guarantees for it. To alleviate the problem, researchers have developed methods to either probe the model's gradient [46, 50, 52, 67], as used in a recent AI-for-code paper [44], or to fit explainable linear models around a small local region of the model's prediction boundary [16, 36, 43]. Explanation methods have also been created for graphical neural networks. GAT [57] and CGCN [58] use attention mechanism to attribute edge importance, while GnnExplainer [61] tries to mask out irrelevant edges and node features while maintaining max mutual information between inputs and outputs. To explain models via concepts and prototypes, [31] proposes a method to learn the similarity between inputs and a small set of prototypes during training via an auto-encoder structure. Our approach is complementary to these approaches as it treats models as black boxes and generates the *precise minimal representation* of an input that a model absolutely requires to arrive at its prediction. The added benefit is that P2IM can finalize the explanation without approximating the proper threshold for ranked features, which is considerably more convenient and actionable for end-users. In addition, the emitted minimal sequence is valid compilable code and thus can be independently cross-verified with existing source code analyzers.

Finally, with regard to deep learning models' robustness and reliability research, metrics have been proposed for distance ratio among samples in the same class and in all other classes [24]. This is based on the intuition that, for a reliable model, inputs in the same class should be closer to each other in the model's latent space, as compared to inputs in the other class. Extensive research has also contributed to constructing sophisticated attack/defense methods and evaluating accuracy under $\ell_p$-norms bounded adversarial perturbations [11, 15, 29, 37, 39, 49]. Different from current metrics, we directly probe the existence of true signal inside the models' 1-minimal representations. To the best of our knowledge, our metric *SAR* is the first of its kind to evaluate deep learning models' signal awareness in this domain.

## 7 CONCLUSION

In this paper, we present a prediction-preserving input minimization approach called P2IM to evaluate and compare the signal awareness of AI-for-code models. In particular, P2IM systematically reduces a program sample to a minimal snippet which a model needs to arrive at and stick to its original vulnerable prediction. By checking if the minimal snippet has the same vulnerability as the original sample, P2IM measures the model's reliance on incorrect signals. We apply P2IM on three state-of-the-art neural network models across multiple datasets, and measure their signal awareness using a new metric we propose – Signal-aware Recall (SAR). The results show a sharp performance drop, which suggests the models are presumably learning a lot of noises or dataset nuances, as opposed to capturing vulnerability-related signals. SAR augments the traditional measures of model performance with a new metric to measure task-relevant learning, which can more fairly compare and guide improvements across model evolutions.